\documentclass{INTERSPEECH2023}
\usepackage{amsmath}
\usepackage{graphicx}
\usepackage{subcaption}
\usepackage{mwe}
\usepackage{mathtools}
\usepackage{multirow}
\usepackage{float}
\usepackage[margins]{trackchanges}
\addeditor{NK}

\DeclarePairedDelimiter\ceil{\lceil}{\rceil}

\interspeechcameraready

\title{Weakly-supervised forced alignment of disfluent speech using phoneme-level modeling}
\name{Theodoros Kouzelis$^1$, Georgios Paraskevopoulos,$^{1,2}$, Athanasios Katsamanis$^1$, Vassilis Katsouros$^1$}
\address{
  $^1$ Institute for Language and Speech Processing, Athena Research Center, \\
  $^2$School of ECE, National Technical University of Athens, Athens, Greece}
\email{theodoros.kouzelis@athenarc.gr, g.paraskevopoulos@athenarc.gr, nkatsam@athenarc.gr, vsk@athenarc.gr}

\begin{document}

\maketitle
 
\begin{abstract}

The study of speech disorders can benefit greatly from time-aligned data. However, audio-text mismatches in disfluent speech cause rapid performance degradation for modern speech aligners, hindering the use of automatic approaches.
In this work, we propose a simple and effective modification of alignment graph construction of CTC-based models using Weighted Finite State Transducers. The proposed weakly-supervised approach alleviates the need for verbatim transcription of speech disfluencies for forced alignment. During the graph construction, we allow the modeling of common speech disfluencies, i.e. repetitions and omissions. Further, we show that by assessing the degree of audio-text mismatch through the use of Oracle Error Rate, our method can be effectively used in the wild.
Our evaluation on a corrupted version of the TIMIT test set and the UCLASS dataset shows significant improvements, particularly for recall, achieving a 23-25\% relative improvement over our baselines.

\end{abstract}
\noindent\textbf{Index Terms}: fluency disorder, phonetic
alignment, WFST

\section{Introduction }

Speech disfluencies are common for both children and adults \cite{stutter_rate}. Stuttering when reading aloud or speaking spontaneously is an example of a disfluency that can have a profound impact on an individual's ability to communicate effectively \cite{comm}. 
Stuttering typically begins in childhood \cite{child_stutt} and manifests itself as interruptions in the flow of speech such as repeating words or sounds and committing false starts.
Diagnosis and assessment of such disfluencies have traditionally been done by clinicians, who manually count stuttering events after having transcribed and aligned text and audio of recorded sessions to classify disfluencies \cite{uclass}. Disfluencies in both spontaneous speech and aloud reading are not prevalent only in the speech of individuals with speech disorders. In \cite{Shriberg1994PreliminariesTA}, Shriberg demonstrates that there is a $50\%$ probability for a sentence of $10–13$ words to include a disfluency and that the probability increases with sentence length.
\begin{figure}[t]
  \centering
  \caption{An overview of our proposed approach. Given an approximate transcription, the disfluent verbatim utterance is aligned with the corresponding sections of the audio.}
  \includegraphics[scale=0.19]{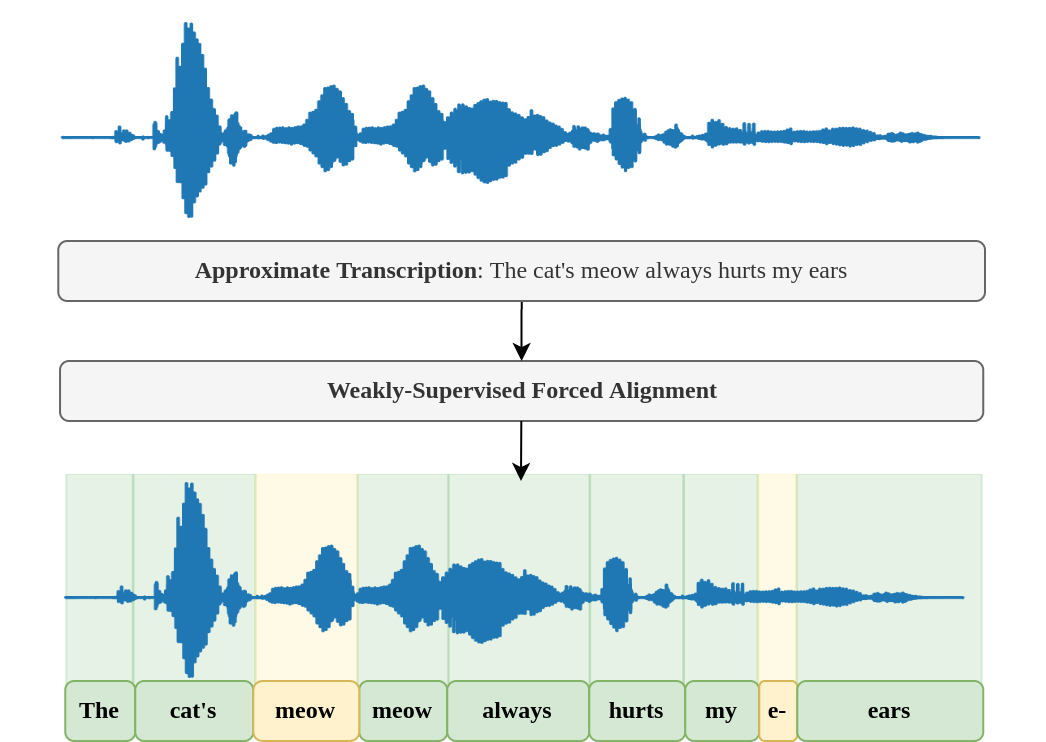}
  \label{fig:overviw}
\vspace{-15 pt}
\end{figure}

In studies of stuttering or other disfluencies, the analysis of large corpora is essential due to the rich inherent variability in speech production. However, the traditional method of manually annotating and aligning words and phonemes is an extremely time-consuming process and can take up to 
130 times real-time \cite{manual}.
Automatic forced alignment, which detects time boundaries for each phoneme or word of already transcribed speech, can replace or speed up manual alignment depending on the required level of accuracy \cite{speedup}.

 \begin{figure*}[!htp]
        \centering
        \caption{The composing parts of the alignment graph $\mathcal{A}$. (a), (b) present the linear FSA and the Modified FSA constructed on the sentence “don’t ask” with CMU phonetic transcription [D AA N T AE S K]. Part-word repetition arcs (PW) are illustrated in purple, word repetition arcs (W) in blue, and word deletion arcs (D) in red.  (b), (d) are an example Emissions Graph and the standard CTC Topology Graph constructed over the CMU phone set. For clarity, we illustrate the CTC topology over two phones \{AA, AE\}.}
        \begin{subfigure}[]{0.17\textwidth}
            \centering
            \includegraphics[width=270pt]{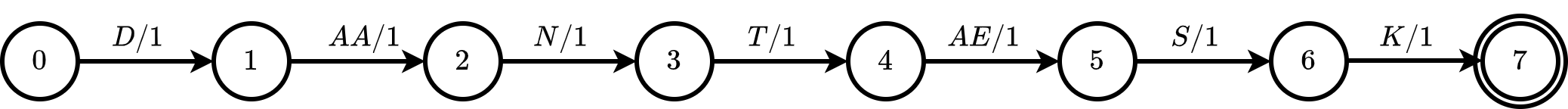}
            \caption[Linear FSA]%
            {{\small Linear FSA $\mathcal{Y}$}}    
            \label{fig:linear}
        \end{subfigure}
        \hfill
        \begin{subfigure}[]{0.44\textwidth}  
            \centering 
            \hspace*{0.5 cm}
            \includegraphics[width=\textwidth]{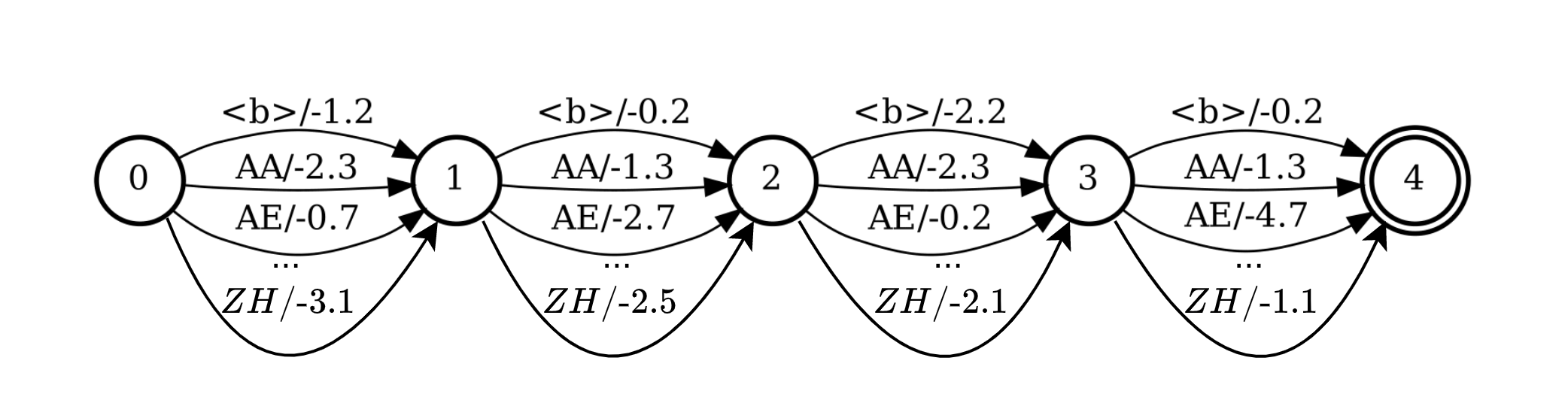}
            \caption[]%
            {{\small Emissions Graph $\mathcal{E}_x$}}    
            \label{fig:emissions}
        \end{subfigure}
        \vskip -27pt 
        \vskip\baselineskip
        \begin{subfigure}[]{0.17\textwidth}   
            \centering 
            \includegraphics[width=270pt]{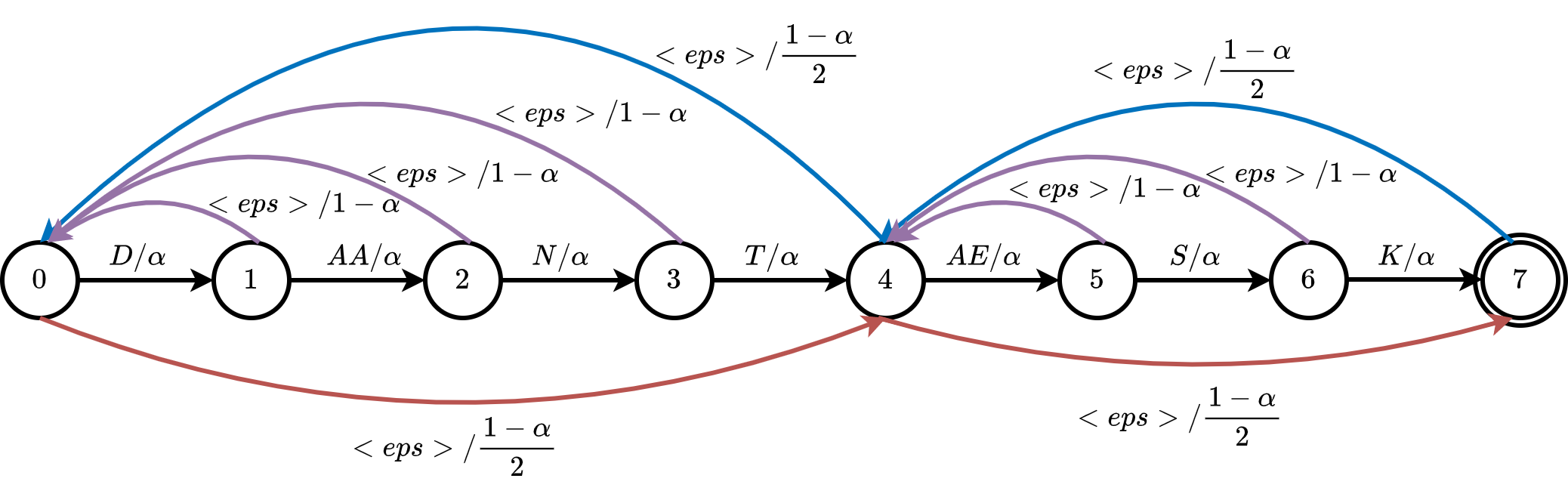}
            \caption[]%
            {{\small Modified FSA $\mathcal{Y}_M$}}    
            \label{fig:modified}
        \end{subfigure}
        \hfill
        \begin{subfigure}[]{0.33\textwidth} 
            
            \centering 
            \hspace*{-0.5cm}
            \includegraphics[width=\textwidth]{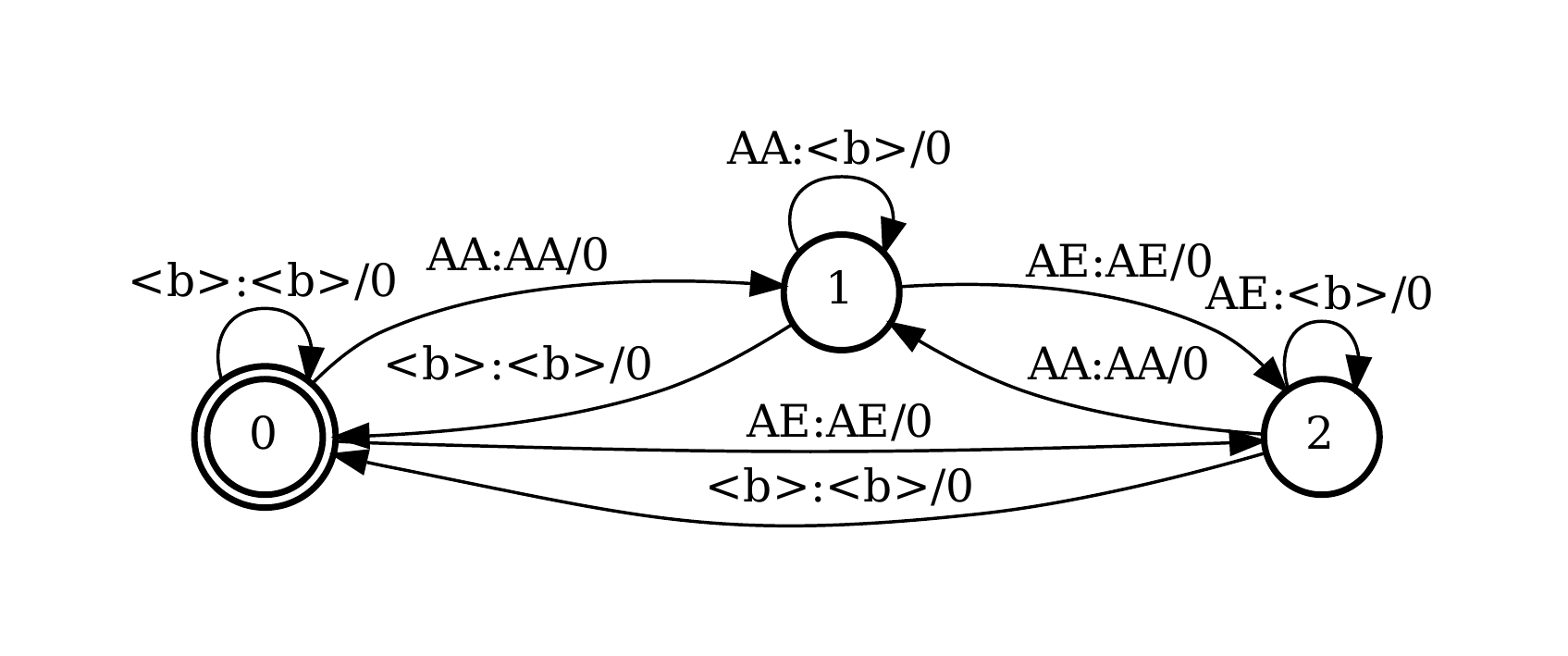}
            \caption[]%
            {{\small CTC Topology Graph $\mathcal{T}$}}    
            \label{fig:topo}
        \end{subfigure}
        {\small } 
        \label{fig:mean and std of nets}
        \vspace{-10 pt}

    \end{figure*}

Manual transcription processes, however, typically correct or remove any disfluency events, resulting in transcripts that are considerably cleaner than the actual speech \cite{switchboard_dis}. Further, when speech contains disfluencies, it is necessary to manually transcribe any corresponding events before the alignment since the performance of conventional forced alignment approaches can be significantly affected by mismatches between speech and text \cite{sail}. 
Therefore, it is crucial for robust speech-to-text alignment to be able to detect disfluencies and align them accurately with the appropriate parts of the utterance.

Conventionally, speech-to-text alignment is performed using Hidden
Markov Models - Gaussian Mixture Model (HMM-GMM) systems by application of the standard Viterbi-based forced alignment. However, very long audio and potential mismatches between speech and text pose a challenge to these systems as they are susceptible to error propagation. In the literature these issues are usually mitigated by iterative approaches \cite{sail, moreno, Hazen}.
In \cite{Hazen}, the alignment and correction of approximate transcriptions are addressed with an iterative approach and the use of a modified lattice that allows word insertions and deletions. A similar approach with a grammar-based decoder has been used for the alignment of dysarthric speech \cite{dysarth}.
Prior research has explored the effectiveness of using an augmented decoding grammar based on the original transcription to identify disfluencies, including word and syllable repetitions, in both stuttering detection \cite{stutter} and disfluency detection during oral reading \cite{reading, tutor}. However, these studies have not addressed the issue of performance degradation of forced alignment when faced with untranscribed disfluencies.

While HMM-GMM based techniques have been thoroughly researched for the alignment of speech and text, recent works are exploring the use of neural networks with promising results \cite{chasiu,ali_cor, ctc_ali, ali_comp}. 
In most recent neural forced aligners the final alignment is extracted from the emission probabilities by Dynamic Time Wrapping (DTW) \cite{chasiu} or by 
following the procedure in \cite{ctc_ali} if the blank token of Connectionist Temporal Classification (CTC) training is present i.e., cumulative
score, blank distribution, and beam search decoding \cite{teytaut22_interspeech}. These methods limit the modifiability made possible by decoding with Weighted Finite State Transducers (WFSTs). WFST representations enable the efficient incorporation of lexicons and language models into CTC decoding \cite{essen}, memory-efficient variations of the standard CTC \cite{ctc_var}, and even training with partially labeled data \cite{star}.

In this work, we investigate the performance degradation of state-of-the-art forced aligners under untranscribed disfluencies. We demonstrate that given the emission probabilities of a frame classification or CTC-based model, the forced alignment can be achieved with WFST operations. Building on this, we propose a weakly-supervised forced alignment that aims at aligning the full verbatim disfluent utterance without the need for manual transcription. An overview of our proposed system is presented in Figure 1.
To address an in the wild scenario, where the severity of audio-text mismatch is unknown we constrain the degrees of freedom of the weakly-supervised forced alignment leveraging Oracle Error Rate \cite{farfield}.
To test our method we employ a \textit{state-of-the-art} neural forced aligner \cite{chasiu} that is based on Wav2Vec2 \cite{wav2vec}. We evaluate our approach on the UCLASS dataset \cite{uclass} and a corrupted version of the TIMIT test set with synthesized disfluencies. 

Our key contributions can be summarized as follows: (a) We address the performance degradation of forced alignment under untranscribed disfluencies, (b) we introduce a weakly supervised forced alignment that is applicable to modern CTC-based neural forced aligners, 
(c) we construct and provide \mbox{DisfluenTIMIT}, a testbed for the evaluation of forced alignment under untranscribed disfluencies which we hope
 will inspire further research on the topic. Our code is available as open-source\footnote{https://github.com/zelaki/WSFA/}.

\section{Method}
\label{sec:method}
For the methods described in this Section, we assume: (a) an input sequence $X = [x_0, ..., x_{T-1}]$ corresponding to the speech signal, (b) a phonetic transcription $Y= [y_0, ..., y_{N-1}]$ , $y_n \in \mathcal{P} $, where $\mathcal{P}$ is the phonetic vocabulary, and (c) a network that given $X$  generates frame-based phone posteriors in the log-domain $\log P(p_t \vert X)$ with $p_t \in \mathcal{P}$ and $t \in T$.

\noindent\textbf{Weakly supervised forced alignment with WFSTs}:
Given the phonetic transcription $Y$, a label graph $\mathcal{L}_{Y}$ that represents all the valid alignment paths can be constructed.
Further, using the phone posterior log probabilities, an emission graph $\mathcal{E}_x$ (Figure \ref{fig:emissions}) can be created, where arc weights correspond to $\log P(p_t \vert X)$.
The alignment graph is derived by composing the emission and label graphs, $\mathcal{A} = \mathcal{L}_{Y} \circ \mathcal{E}_x$. The shortest path in $\mathcal{A}$ is the final frame-level alignment.  
 The label graph can be further deconstructed to its composing parts $\mathcal{L}_{Y} = \mathcal{T} \circ \mathcal{Y}$, a topology graph $\mathcal{T}$ (Figure \ref{fig:topo}) and a linear Finite State Acceptor (FSA) $\mathcal{Y}$ (Figure \ref{fig:linear}),  that accepts only the input phonetic transcription $Y$. Thus the alignment graph can be written as:
 \begin{align}
     \mathcal{A} = \mathcal{T} \circ \mathcal{Y} \circ \mathcal{E}_x
 \end{align}


To enable the automated recognition and alignment of disfluencies, we suggest a straightforward adaptation to the linear FSA. 
Given an approximate transcription, a Modified FSA $\mathcal{Y}_M$ incorporating non-consuming $<eps>$ arcs is constructed. 
Figure \ref{fig:modified} shows an example Modified FSA for the sentence “don't ask” [\textit{D AA N T AE S K}]. 
Following the horizontal arcs, the original transcription is obtained.
Traversing the blue arcs allows both word and phrase repetitions, such as “don't ask don't ask”. Red arcs allow the deletion of a word or phrase. Omissions of words are frequently observed during oral reading, particularly in children \cite{deletions}. Part-word repetitions
are allowed by the purple arcs. Unlike the word lattices described in \cite{stutter} and \cite{reading}, our approach entails modeling of part-word repetitions at the phoneme level rather than the syllable level, alleviating the need for manual segmentation of words into syllables. Intra-word repetitions are not modeled since they are not as common in disfluent speech \cite{stutter}. 

The arc weights of $\mathcal{Y}_M$ are constrained by a prior probability $\alpha$ which restraints excessive changes to the input phonetic transcription. For each state, there is a probability of $\alpha$ assigned to traversing a horizontal arc that follows the path outlined by the transcription. Conversely, the probability of traversing a non-consuming $<eps>$ arc is $1 - \alpha$. In the case of the start and end word states, the probability is equally divided between the deletion (red) and the repetition (blue) arc. When $\alpha = 1$ the Modified FSA degenerates to the linear FSA. 
 As noted by prior studies \cite{stutter, reading,audiobooks} the weights of $\mathcal{Y}_M$ are effective when $\alpha \geq 0.9$ as the log probabilities of graph $\mathcal{E}_x$ are of larger orders of magnitude. Thus we will denote $\alpha = 1 - 10^{- \beta}$ with $\beta \in [0,+\inf)$ and consider values of $\beta$.
\\ 
\noindent\textbf{Adaptively choosing a value for $\beta$}:
The modified FSA adds degrees of freedom in the forced alignment process allowing the recognition of untranscribed events, but also the possibility of false alarms. Thus, the value of $\beta$ must balance this trade-off.
In previous works \cite{stutter, reading, audiobooks}, the arc weights are empirically decided to fit the target data.
But in a real word scenario, the mismatch between audio and transcription as well as the severity of speech disfluencies are unknown.  So, the value of $\beta$ should be proportional to the mismatch between the spoken utterance and the transcription. A simple method to quantify this mismatch is Oracle Error Rate (OER) \cite{farfield}.  
A low OER indicates that the transcription of the utterance is mostly correct. We calculate OER as in \cite{farfield,oer}, with a biased bi-gram language model.

Figure 3 demonstrates that utilizing OER is an effective means of distinguishing between clean and disfluent samples. Building on this, we derive an intuitive formula to condition $\beta$, and consequently the degrees of freedom of the weakly-supervised forced alignment, on OER. For a sampled audio-transcription pair $S$  we set $\beta_S = 10^{(1-O_S)} $, where $O_S$ is the OER of sample $S$. Since  OER can be greater than 1, we explicitly restrict $O_S \in [0,1]$ and, thus, $\beta \in [1,10]$. 

 \begin{figure}[t]
  \centering
  \caption{Oracle Error Rates for TIMIT and DisfluenTIMIT test sets. DisfluenTIMIT is presented in Section \ref{sec:data}.}
  \includegraphics[scale=0.4]{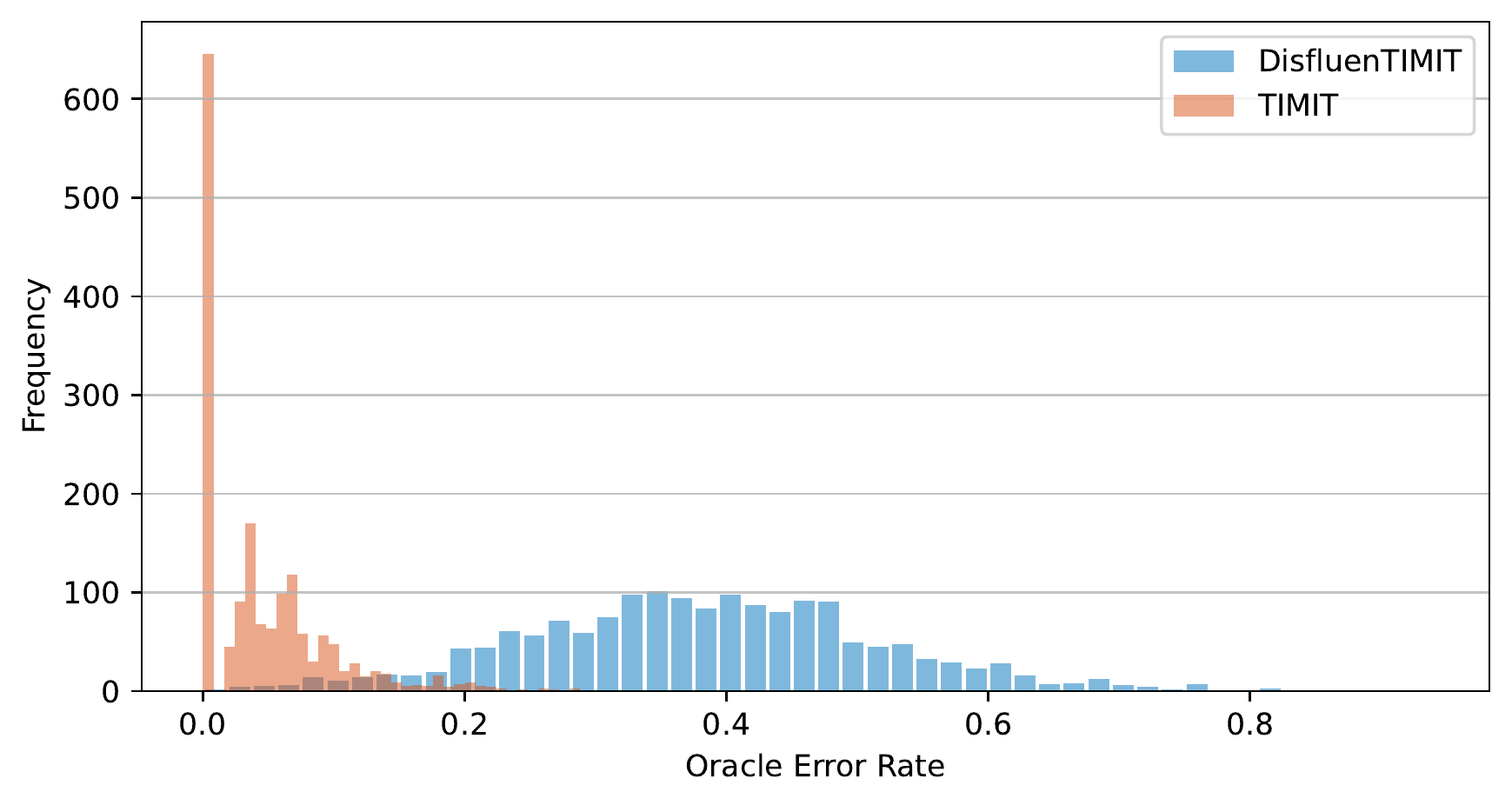}
  
  \vspace{-5mm}
  \label{fig:overviw}
\end{figure}

\begin{table*}[!htb]
\caption{Evaluation results on UCLASS and DisfluenTIMIT. WFST and WFST\_WS denote the standard and weakly-supervised force alignment presented in Section \ref{sec:method}.}
\centering
\label{tab:main}
\scalebox{0.87}{
\begin{tabular}{|c|c|cc|cc|cc|cc|cc|}
\hline
Dataset                        & Method & P    & $\text{P}_{\downarrow \%}$ & R    & $\text{R}_{\downarrow \%}$ & F1   & $\text{F1}_{\downarrow \%}$ & R-val & $\text{R-val}_{\downarrow \%}$ & Overlap & $\text{Overlap}_{\downarrow \%}$ \\ \hline
\multirow{4}{*}{DisfluenTIMIT} & \texttt{MFA}    & $0.56$ & $14.9$            & $0.48$ & $29.4$            & $0.51$ & $23.4$            & $0.58$  & $17.9$            & $0.50$    & $24.4$            \\
                        & \texttt{DTW}       & $0.59$ & $9.0$  & $0.47$ & $22.4$ & $0.52$ & $17.4$ & $0.59$  & $14.1$ & $0.57$ & $15.4$ \\
                        & \texttt{WFST}      & $0.57$ & $12.3$ & $0.47$ & $21.9$ & $0.51$ & $18.3$ & $0.584$ & $14.4$ & $0.57$ & $15.3$ \\
                        & \texttt{WFST\_WS(Ours)} & $\textbf{0.61}$ & $\textbf{5.8}$  & $\textbf{0.60}$ & $\textbf{1.7}$  & $\textbf{0.60}$ & $\textbf{3.8}$  & $\textbf{0.66}$  & $\textbf{3.4}$  & $\textbf{0.67}$ & $\textbf{1.1}$  \\ \hline
\multirow{4}{*}{UCLASS} & \texttt{MFA}               & $0.36$ & $28.4$ & $0.37$ & $33.1$ & $0.36$ & $30.8$ & $0.42$  & $23.4$ & -    & -    \\
                        & \texttt{DTW}       & $0.54$ & $\textbf{1.67}$ & $0.52$ & $4.8$  & $0.53$ & $3.3$  & $0.57$  & $2.36$ & -    & -    \\
                        & \texttt{WFST}      & $0.55$ & $2.2$  & $0.52$ & $6.0$  & $0.54$ & $4.1$  & $\textbf{0.58}$  & $3.1$  & -    & -    \\
                        & \texttt{WFST\_WS(Ours)}       & $\textbf{0.56}$ & $1.9$  & $\textbf{0.54}$ & $\textbf{3.8}$  & $\textbf{0.54}$ & $\textbf{2.9}$  & $\textbf{0.58}$  & $\textbf{2.3}$  & -    & -    \\ \hline
\end{tabular}
}
\end{table*}

\section{Data}
\label{sec:data}
The scarcity of speech disfluency datasets and the need for manual time-aligned data pose significant challenges for evaluating speech alignment methods. To overcome these challenges, we create the DisfluenTimit dataset by introducing speech disfluencies into the standard Timit test set. Table \ref{tab:datasets} displays the disfluency types and their percentage of occurrence calculated over the total number of words in the two datasets.

\noindent\textbf{UCLASS}
 is a commonly used dataset in disfluency-related machine learning studies, consisting of 457 audio recordings of children with known stuttering issues. UCLASS Release Two contains five recordings with manually aligned transcriptions in orthographic format, where only the onset of each word or disfluency is labeled. Since disfluencies are flagged we are able to extract the approximate transcription for our experiments.

\noindent\textbf{DisfluenTIMIT}
is a synthetic dataset that includes various types of disfluencies, including part-word repetitions, word repetitions, phrase repetitions, and word deletions. To generate these disfluencies, we followed a very similar approach to \cite{libristutter}, sampling and copying the audio or removing it in the case of deletions. For an audio file with $n$ words, we randomly sample a disfluency percentage $p \in [0.1, 0.2, 0.3]$ and insert $\ceil{p \cdot n}$ random disfluencies of types $d \in [PW, W, PH, D]$. The disfluencies are labeled accordingly, and the ground truth disfluent phonetic transcription is acquired.
\begin{table}[H]
\small
\caption{\label{tab:datasets}Types of disfluencies in the samples from UCLASS and DisfluenTimit. The $3^d$ and $4^{th}$ row present average percentage of each type.}
\centering
\scalebox{0.9}{
\begin{tabular}{|c|c|c|c|}
\hline
Label & Disfluency Type       & DisfluenTimit & UCLASS \\ \hline
PW    & Part-word repetitions & $4.7\%$         & $0.4\%$  \\ \hline
W     & Word repetitions      & $4.8\%$         & $7.0\%$  \\ \hline
PH    & Phrase repetitions    & $4.7\%$         & $0.7\%$  \\ \hline
D     & Word deletions        & $4.7\%$         & $0\%$   \\ \hline
I     & Injections            & $0\%$           & $2.3\%$  \\ \hline
\end{tabular}
}
\vspace{-5 pt}
\end{table}
More specifically, to generate word repetitions, we repeat a randomly selected word one, two, or three times before the original utterance. For phrase repetitions, we repeat a phrase consisting of two to three words. For part-word repetitions, we copy and append a random number of phones before the selected word. For word deletions, we delete one, two, or three consecutive words from the audio. To ensure a smoother transition between the added audio and the existing clip, we use interpolation as described in \cite{libristutter}. The original TIMIT 61 phones are collapsed into the 39 CMU phone set as in \cite{chasiu}.



\section{Experimental Setup}
 To evaluate the methods described in Section 3 we utilized a pre-trained \textit{state-of-the-art} frame classification model called \texttt{W2V2-FC}\footnote{https://huggingface.co/charsiu/en\_w2v2\_fs\_10ms}\cite{chasiu}  that has been trained on Librispeech \cite{librispeech}. We compare our proposed method against two strong baselines: \texttt{Montreal Forced Aligner (MFA)} and \texttt{W2V2-FC} with \texttt{DTW} as in \cite{chasiu}. All WFST operations described in Section 3 are executed using the k2 framework\footnote{https://github.com/k2-fsa/k2}.
For each audio, we extract the emission probabilities using \texttt{W2V2-FC}, resulting in a 2D tensor $E \in \mathbb{R^{N \times T}}$ where  T is the number of frames and N=42, including the 39 CMU phones and special tokens such as [SIL] for silence, [UNK] for unknown phones, and [PAD] for padding.
 As samples in the UCLASS dataset are longer, we divide the audio into 10-second chunks, feed them through the model, and concatenate the output emission probabilities in the time dimension.
 The [SIL] token and the \textit{blank} token $<b>$ of the topology graph $\mathcal{T}$ are mapped in the same token id, as \texttt{W2V2-FC} has been fine-tuned with the cross-entropy loss instead of CTC, and  thus the blank token is absent. In all our experiments the transition weights of the Modified FSA in the weakly-supervised approach are adaptively chosen for each sample as described in Section \ref{sec:method}.

 We evaluate the alignment accuracy using precision, recall, F1 score, and R-value \cite{rval}. A hit is counted for each predicted phone boundary that falls within a tolerance of $\tau$ and matches the predicted phone.  For DisfluenTIMIT, we only evaluate the phone onsets since each boundary marks the onset and offset of consecutive phones, while for UCLASS, we evaluate only the word onsets. We use a tolerance of 40ms and 100ms for DisfluenTIMIT and UCLASS respectively. Additionally, we measure the percentage of accurately predicted frame labels at a 10ms timescale for DisfluenTIMIT (Overlap metric in Table \ref{tab:main}).
 The relative performance reduction for each metric $M$ is calculated as:
\begin{equation}
    M_{\downarrow \%} = \frac{M_{verbatim} - M_{approximate}}{M_{verbatim}} \times 100
\end{equation}

\begin{table}[H]
\caption{The effect of word repetition $W$, word deletions $D$ and part-word repetition $PW$ arcs on performance metrics for DisfluenTIMIT dataset. Minus (-) denotes without.}
\label{tab:arcs}
\centering
\scalebox{0.9}{
\begin{tabular}{c|c|c|c|c|c}
Method & P & R & F1 & R-val & Overlap \\ \hline
\texttt{WFST\_WS} & $0.61$ & $0.60$ & $0.60$ & $0.66$ & $0.67$ \\
\texttt{-PW} & $0.59$ & $0.58$ & $0.58$ & $0.64$ & $0.65$ \\
\texttt{-D} & $0.59$ & $0.59$ & $0.58$ & $0.63$ & $0.66$ \\
\texttt{-W} & $0.59$ & $0.47$ & $0.52$ & $0.60$ & $0.57$ \\
\texttt{-W-D} & $0.57$ & $0.47$ & $0.51$ & $0.59$ & $0.56$ \\
\texttt{-W-PW} & $0.59$ & $0.47$ & $0.52$ & $0.59$ & $0.57$ \\
\texttt{-PW-D} & $0.57$ & $0.58$ & $0.58$ & $0.62$ & $0.64$
\end{tabular}
}
\end{table}

Where, $M_{approximate}$ and $M_{verbatim}$ denote the performance metrics calculated with untranscribed and transcribed disfluencies, respectively.

To evaluate the effectiveness of the adaptive approach presented in Section \ref{sec:method} for adjusting the Modified FSA weights, we conduct an experiment that simulates a real-world scenario where a combination of disfluent and clear speech samples are encountered without prior knowledge of the speech-text mismatch. We create a mixed test set, referred to as MixTIMIT, by randomly selecting the disfluent or fluent version of a sample from the DisfluenTIMIT and TIMIT test sets, respectively.
 \section{Results and Discussion}
Table \ref{tab:main} displays the alignment performance of all methods on both datasets.
Weakly-supervised forced alignment with WFSTs demonstrates superior performance across all metrics, with the smallest performance reduction when disfluencies are not transcribed. However, it is worth noting that while recall and frame overlap are significantly improved, precision is only marginally affected due to the presence of false alarms.
Interestingly, \texttt{DTW} outperforms \texttt{MFA} in both datasets indicating that it is less affected by mismatches between audio and transcription. 

\begin{figure}[H]

  \centering
  \includegraphics[width=0.83\columnwidth]{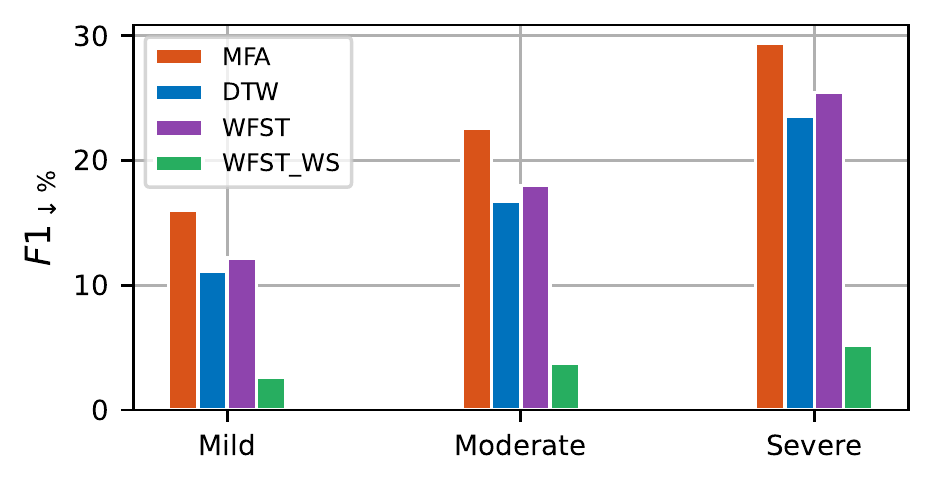}
  \caption{Relative performance reduction of F1 metric under mild, moderate and severe untranscribed disfluencies.}
  \label{fig:severity}
  \vspace{-5 pt}
\end{figure}

To investigate the impact of the number of untranscribed disfluencies on the performance, we divide uniformly DisfluenTIMIT into three categories based on the percentage of disfluencies in relation to the total number of words. 
The results are presented in Figure \ref{fig:severity}, which clearly demonstrates that our proposed approach is robust to untranscribed disfluencies.
In contrast, all other methods and especially \texttt{MFA}, experience significant performance degradation, particularly in severe cases where the percentage of disfluencies is high. 

\noindent\textbf{Additional arcs}: In Table \ref{tab:arcs} we present the performance gains associated with each of the additional arcs to gain insight into the relative importance of these components. Our results indicate that omitting repetition arcs $W$ has the most significant impact on alignment performance, suggesting that untranscribed word or phrase repetitions are a primary cause of performance degradation of forced alignment.

\noindent\textbf{Effects of adaptive $\beta$:}
Using the MixTIMIT test set we compared the impact of adaptively choosing $\beta$ for each sample versus globally setting $\beta$ to an a priori value for all test samples. The results are presented in Table \ref{tab:adaptive}. While the performance isn't improved substantially, our adaptive approach succeeded at mitigating the necessity of prior knowledge about the severity of untranscribed speech disfluencies in target data or an extensive grid search.

\begin{table}[H]
\caption{Global (G) vs adaptive (A) $\beta$ ablation on MixTIMIT.}
\centering
\label{tab:adaptive}
\scalebox{0.9}{
\begin{tabular}{c|c|c|c|c|c}
$\beta$  & P    & R    & F1   & R-val & Overlap \\ \hline
1 (G)    & $0.69$ & $0.67$ & $0.68$ & $0.72$  & $0.63$    \\
10 (G)  & $0.69$ & $0.67$ & $0.68$ & $0.73$  & $0.63$    \\
100 (G) & $0.68$ & $0.65$ & $0.66$ & $0.71$  & $0.62$    \\
1000 (G) & $0.67$ & $0.61$ & $0.64$ & $0.69$  & $0.58$   \\
$10^{(1-O_S)}$ (A) & $0.70$ & $0.68$ & $0.69$ & $0.74$  & $0.63$    

\end{tabular}
}
\end{table}


\section{Conclusions and Future Work}
In this paper, we have explored the performance decrease of state-of-the-art forced aligners under untranscribed disfluencies. We proposed a weakly-supervised forced alignment, by a straightforward modification of the alignment graph construction of CTC-based models, utilizing WFSTs. 
Our method enables the alignment of common disfluencies without the need for manual transcription while incurring minimal performance decrease when disfluencies are not transcribed. Utilizing OER to quantify the severity of untranscribed disfluencies before alignment enables the effective application of our approach in the wild.
In feature work, we plan to further explore the use of WFSTs to model pronunciation variants and other disfluency types such as injections, and also make the Modified FSA's weights learnable through Expectation Maximization.



\bibliographystyle{IEEEtran}
\bibliography{template}

\end{document}